\documentclass[mathematics,article,accept,moreauthors,pdftex]{mdpi}
\usepackage{algorithm}
\usepackage{algpseudocode}
\usepackage{doi}

\algdef{SE}[DOWHILE]{Do}{doWhile}{\algorithmicdo}[1]{\algorithmicwhile\ #1}

\newcommand{\eps}{\varepsilon}
\newcommand{\nulmt}{\textbf{0}}

\newcommand{\reff}[1]{(\ref{#1})}
\newcommand{\tr}{^{\scriptsize\mbox{\textup{T}}}}
\newcommand{\prodser}[3]{\disp\sum_{j=0}^{#1}\langle #3Q_{#2}#3\lam_j,#3\lam_{#1-j}\rangle}
\newcommand{\disp}{\displaystyle}
\newcommand{\nl}{\medskip\\}
\newcommand{\lam}{\Upsilon}

\firstpage{1} 
\pubvolume{10}
\issuenum{8}
\articlenumber{1207}
\pubyear{2022}
\copyrightyear{2022}
\datereceived{3 March 2022} 
\dateaccepted{5 April 2022} 
\datepublished{7 April 2022} 
\hreflink{https://doi.org/10.3390/math10081207}

\Title{On a High-Precision Method for Studying Attractors of Dynamical Systems and Systems of Explosive Type}

\TitleCitation{On a High-Precision Method for Studying Attractors of Dynamical Systems and Systems of Explosive Type}

\Author{Alexander N. Pchelintsev \orcidA{}}

\AuthorNames{Alexander N. Pchelintsev}

\AuthorCitation{Pchelintsev, A.N.}

\address[1]{%
Tambov State Technical University, Department of Higher Mathematics,
ul. Sovetskaya 106, 392000  Tambov,  Russia; pchelintsev.an@yandex.ru}

\abstract{The author of this article considers a numerical method that uses high-precision calculations to construct approximations to attractors of dynamical systems of chaotic type with a quadratic right-hand side, as well as to find the vertical asymptotes of solutions of systems of explosive type.
A special case of such systems is the population explosion model. A theorem on the existence of asymptotes is proved. The extension of the numerical method for piecewise smooth systems is described using the Chua system as an example, as well as systems with hysteresis.}



\begin{document}
\end{paracol}

\section{Introduction}
\label{sec:1}

Let us consider a dynamical system with quadratic nonlinearities of the form
\equation
 \dot{X}=B_0+B_1X+\varphi(X),\label{polsys}
\endequation
where $X(t)=\left[x_1(t)\:\ldots\:x_n(t)\right]\tr$ is a vector function of time $t$ with values in space $\mathbb R^n$, $B_0\in\mathbb R^n$ is a given column vector,
$$
 \varphi(X)=\left[\varphi_1(X)\:\ldots\:\varphi_n(X)\right]\tr,
$$
$\varphi_p(X)=\langle Q_p X,X \rangle$, $B_1$ and $Q_p$ ($p=1,\ldots,n$) are the matrices ($n\times n$) of real numbers.

There has been a strong interest in the systems of the form \reff{polsys} in scientific literature starting somewhere in the middle 20th century.
The most popular object of study from the class of systems \reff{polsys} is the Lorenz system \cite{Lorenz}:
\equation
 \left\{\begin{array}{l}
 \dot{x}_1=\sigma(x_2-x_1),\nl
 \dot{x}_2=rx_1-x_2-x_1x_3,\nl
 \dot{x}_3=x_1x_2-bx_3.
 \end{array}\right.\label{lorenzsys}
\endequation
For this system, the matrices have the form:
$$
 \begin{array}{c}
 B_0=\nulmt,\:\:\:
 B_1=\left[
 \begin{array}{ccc}
 -\sigma & \sigma & 0\nl
 r & -1 & 0\nl
 0 & 0 & -b
 \end{array}
 \right],\:\:\:
 Q_1=\nulmt,\nl
 Q_2=\left[
 \begin{array}{rrr}
 0 & 0 & -1\nl
 0 & 0 & 0\nl
 0 & 0 & 0
 \end{array}
 \right],\:\:\:
 Q_3=\left[
 \begin{array}{rrr}
 0 & 1 & 0\nl
 0 & 0 & 0\nl
 0 & 0 & 0
 \end{array}
 \right].
 \end{array}
$$

Such systems are of interest to researchers because they are systems with energy dissipation. 
There is compression of the volume of the phase space in them, and all trajectories are uniformly bounded. Thus, there is an attracting set called the attractor of the dynamical system to which all solutions tend at $t\rightarrow\infty$.
Consequently, the attractors of dissipative systems determine their behavior over long time intervals. However, for many systems of the form \reff{polsys}, the limit solutions are unstable. Therefore,
correct numerical integration requires the high-precision methods that make it possible to flexibly control the accumulation of calculation errors over a long time interval. For example, with an increase of the order of the Runge--Kutta methods, the complexity of the formulas for calculating the approximate values of the phase coordinates (they have a tree structure) \cite{Hairer,Butcher} grows.

Recently, to simulate of dynamic chaos in systems of the type \reff{polsys}, some researchers have begun to use effective modifications of the power series method. Gibbons in his \mbox{article \cite{Gibbons}} proposed the quick formulas for calculating the coefficients of power series for the main types of right-hand sides of differential equations in \textit{the scalar form}. 
Today this idea was generalized in a recursive procedure (called as automatic differentiation) to compute the values of the derivatives for power series \cite{Rall}. An advantage over the general Taylor series method is that the calculations can be constructed by fast formulas in comparison to the direct symbolic differentiation of right-hand sides of nonlinear ODEs which requires a lot of computer memory for high-precision calculations. The method of power series in \cite{Hashim,Abdulaziz,Sawalha} is applied as the Adomian decomposition method (ADM). The Clean Numerical Simulation (CNS) \cite{Liao} is based on the method of power series at arbitrary-order and used the multiple-precision data, plus a check of solution by means of an additional computation using even smaller numerical noises.

There is the FGBFI numerical method (the firmly grounded backward-forward integration method) described in \cite{LoziPchelintsev,LozPogPchel,Pchelintsev2020,Pchelintsev2021}.
In this articles, the recurrence relations for calculating of the coefficients of expansion of local solutions in a power series are received for any dynamic system with quadratic nonlinearities in \textit{the vector form} (it distinguishes the FGBFI-method from automatic differentiation and CNS). This form of coefficients of the power series is simpler and faster to compute than in ADM (because it does not contain factorials). The author of this article derived the simple formulas of calculating the length of the integration step in \textit{the general form} (it distinguishes the FGBFI-method from CNS). The criteria for checking the accuracy of the approximate chaotic solution are obtained.

This article will consider this method which uses high-precision calculations to construct approximations to unstable solutions of the system \reff{polsys}, as well as its extension to piecewise smooth systems composed of quadratic nonlinearities (for example, the Chua's piecewise linear system).

It is also noteworthy that for certain types of matrices $B_1$ and $Q_p$ and the vector $B_0$, the system \reff{polsys} can have non-extendable solutions of the explosive type. For example, for scalar systems of the form
\equation
 \dot{x}_1=1+x_1^2\label{tan}
\endequation
with a particular solution $x_1(t)=\tan t$ one of the asymptotes is
\equation
 t_{as}=\dfrac{\pi}{2}.\label{astan}
\endequation

The existing classical numerical methods available in the standard implementation of such mathematical packages including Mathcad, MATLAB, Maple, etc., cannot be used to build approximations to such solutions, since the numerical scheme will jump over the asymptote, ''not feeling'' it.
But as will be shown in this article, the FGBFI method allows to get arbitrarily close to the asymptote without skipping the time $t_{as}$, i.e., localizing it.

\section{The FGBFI Method}
\label{sec:2}

The instability of the limiting solutions of the system \reff{polsys}, in fact, makes incorrect to use classical numerical methods over long time intervals.
The limit sets of dynamical systems are constructed on such segments.
We can use high-precision calculations but then we will have rather small integration steps.
The solution to the problem is to apply a modification of the power series method with recurrent calculation of expansion coefficients.

Let us represent the solution of the system \reff{polsys} as a series
\equation
 X(t)=\sum_{i=0}^\infty \lam_i t^i,\label{powser}
\endequation
where the vector $\lam_0$ is determined by the given initial condition. The remaining expansion coefficient vectors \reff{powser} are calculated by the following recursive formulas \cite{LozPogPchel}:
\equation
 \begin{array}{c}
 \Psi_i=\left[\psi_{i,1}\:\:\ldots\:\:\psi_{i,n}\right]\tr,\nl
 \psi_{i,p}=\prodser{i}{p}{},\:\:p=1,\ldots,n,\nl
 \lam_1=B_0+B_1\lam_0+\Psi_0,\nl
 \lam_i=\dfrac{B_1\lam_{i-1}+\Psi_{i-1}}{i}\:\:\:\mbox{for}\:\:i=2,\:3,\:4,\ldots\:.
 \end{array}\label{koef}
\endequation

The convergence region $\left[-\tau(\lam_0),\tau(\lam_0)\right]$ of the series \reff{powser} is limited for many nonlinear systems of the form \reff{polsys}, depends on the initial condition, and can be estimated by the following formulas \cite{LozPogPchel}:
\equation
 \begin{array}{c}
 \disp h_1(\lam_0)=\|\lam_0\|,\:\:\mu=n\cdot\max_{p=1,\ldots,n}\|Q_p\|,\nl
 h_2(\lam_0)=\left\{\begin{array}{l}
 \|B_0\|+(\|B_1\|+2\mu)h_1(\lam_0)+\mu h_1^2(\lam_0),
 \:\:\mbox{if}\:\:h_1>1,\\
 \|B_0\|+\|B_1\|+\mu\:\:\mbox{otherwise},
 \end{array}
 \right.\nl
 \tau(\lam_0)=\dfrac{1}{h_2(\lam_0)+\delta},\:\:t\in\left[-\tau(\lam_0),\tau(\lam_0)\right],
 \end{array}\label{tauc}
\endequation
where $\delta$ is any positive number (can be chosen arbitrarily small). In these formulas, one of the compatible matrix norms $\|\cdot\|_1$ or $\|\cdot\|_{\infty}$ with a vector norm is meant.

The recursive formulas \reff{koef} for calculating the coefficients of the power series expansion were obtained from the substitution of power series into the system \reff{polsys}, as well from the products in Cauchy form. A theorem on the convergence of a power series is proved in the article \cite{LozPogPchel}. The proof is based on the method of mathematical induction. The specified region from formulas \reff{tauc} is obtained as a corollary from this proof.

Next, we describe an algorithm for constructing a trajectory arc on a given time interval $[0,T]$.

In the papers \cite{Pchelintsev2020,Pchelintsev2021}, the numerical method based on the representations \reff{powser}--\reff{tauc} is called the FGBFI method. Next, we describe the algorithm underlying it.

Let $T$ is the given length of the integration interval, $S_a$ is the ball containing the attractor of the system \reff{polsys}. Let us set such a representation of a real number that the machine epsilon
$$
 \eps_m<<\eps_{pw},
$$
where $\eps_{pw}$ is an accuracy of evaluation of the common term of the series \reff{powser}.
Thus, the summation by the Formula \reff{powser} stops at such a value $i=i^*$ when
\equation
 \|\lam_{i^*}\|\cdot|\Delta t|^{i^*}<\eps_{pw},\label{ineq}
\endequation
where $\Delta t$ is an integration step. Note that for the convergence of the series, the value $\Delta t$ must be chosen as
$$
 0<\Delta t\le\tau(\lam_0)\mbox{\:\:\:in forward time}
$$
or
$$
 -\tau(\lam_0)\le\Delta t<0\mbox{\:\:\:in backward time.}
$$

Since the convergence region of the series \reff{powser} is limited, we cannot take $\Delta t=T$ (as mentioned above, the value of $T$ is large). Therefore, the trajectory arc on the time interval $[0,T]$ will consist of the connected arcs for which the series \reff{powser} converges.

Since in the computer simulation we can only operate with a discrete set of phase coordinate vectors (instead of a continuous function $X(t)$), then, by analogy with the Runge--Kutta methods, we will obtain approximate values of phase coordinates through an integration step, only calculations will be carried out at each step using Formulas \reff{powser}--\reff{ineq} for $t=\Delta t=\tau(\lam_0)$.

Note that here the error accumulation at each step is much smaller than for the fixed-order Runge--Kutta methods, since we can quickly enough calculate the missing terms in the sum using the Formulas \reff{powser} and \reff{koef} by reducing the value of $\eps_{pw}$. It can be seen from the formulas \reff{tauc} that with such a numerical integration, the step will be variable.

Let the value $way$ is the direction of integration: $way=1$ for forward time, $way=-1$ for reverse time.

Next, we show an algorithm for obtaining approximate values of phase coordinates.

\begin{algorithm}
	\caption{The FGBFI Method}
	\label{alg:FGBFI}
	\begin{algorithmic}[1]
		\State {\bf Set} the vector $X_0\in S_a$ of initial condition and the value $way$
		\State $ct:=0$ \Comment{The current time}
		\State $ended:=\mbox{{\ttfamily false}}$ \Comment{Are we finishing the algorithm?}
		\State $l:=1$ \Comment{Number of closed interval of series convergence \reff{powser}}
		\State {\bf Calculate} the value of integration step $\Delta t=\tau(X_{l-1})$ using the formulas \reff{tauc}\label{step5}
		\If{$\Delta t>T-ct$}
		\State $\Delta t:=T-ct$
		\State $ct:=T$
		\Else
		\State $ct:=ct+\Delta t$
		\EndIf
		\State $\lam_0:=X_{l-1}$
		\State $X_l:=X_{l-1}$
		\State $p:=1$ \Comment{The product of powers of $t$}
		\State $i:=0$ \Comment{Number of the current member of the series}
		\Do
		\State $i:=i+1$
		\State $p:=p\cdot way\cdot\Delta t$ \Comment{Calculate the current degree of $t=\Delta t$}
		
		                                    \Comment{taking into account the direction of integration; $p$ is negative for odd powers}

		                                    \Comment{when $way=-1$}
		\State {\bf Calculate} the vector $\lam_i$ using the formulas \reff{koef}
		\State $X_l:=X_l+\lam_i\cdot p$ \Comment{Add to $X_l$ the current member of the series}
		\State $L:=\|\lam_i\|\cdot|p|$ \Comment{The right side of inequality \reff{ineq}}
		\doWhile{$L>\eps_{pw}$}
		\State {\bf Print} the value of time $way\cdot ct$ and the vector $X_l$ of the received phase coordinates
		\If{$X_l\notin S_a$}
		\State We moved beyond the ball $S_a$
		\State {\bf Print} ''Decrease the value $\eps_{pw}$ and/or $\eps_{m}$''
		\State $ended:=\mbox{{\ttfamily true}}$
		\EndIf
		\If{$ct=T$}
		\State $ended:=\mbox{{\ttfamily true}}$
		\EndIf
		\If{$ended$}
		\State {\bf Terminate the algorithm}
		\EndIf
		\State $l:=l+1$
		\State {\bf GoTo} Step \ref{step5}
	\end{algorithmic}
\end{algorithm}

The criteria for checking the accuracy of the resulting approximate solution are described in \cite{LoziPchelintsev, LozPogPchel, Pchelintsev2021}. Note that the backward time pass is used in some of them.

\section{The Software Implementation of the FGBFI Method and Numerical Simulations}
\label{sec:3}

For the software implementation of Algorithm \ref{alg:FGBFI} with
$$
 B_0=\nulmt,
$$
and for the study of the limit points of dynamic systems of the form \reff{polsys} for Poisson stability and the calculation of the Lyapunov exponents by the modified Benettin's algorithm \cite{Pchelintsev2020,Pchelintsev2021}, a software in the C++ language for Linux was developed. The source codes are available on GitHub \cite{PchelProg}.

To perform matrix operations in C++, the algorithms and template types of the uBLAS library of the linear algebra collection of the Boost class libraries were used. Note that the uBLAS library gives a large gain in time.

Usually, in calculations, many researchers work with single or double precision subnormal real numbers represented in the IEEE 754 format.
The main disadvantage here is the fixed accuracy of the representation of real numbers which may not allow us to numerically construct approximations to unstable solutions of systems of differential equations over long time intervals.
Therefore, for high-precision calculations, the MPFR C++ library \cite{MPFR} is used.

\textls[-15]{Algorithm \ref{alg:FGBFI} was applied to check the accuracy of \mbox{\textit{the found unstable cycle} \cite{PchelDiff}} in the system \reff{lorenzsys}:}
\equation
 x_1(0)\approx -2.147367631,\:\:
 x_2(0)\approx 2.078048211,\:\:
 x_3(0)= 27.\label{initvalcycle1}
\endequation
The period value is obtained equal to $T_c \approx 1.558652210$.

The initial values \reff{initvalcycle1} were checked on the period in the computer program \cite{PchelProg} that implements the FGBFI method with $\eps_{pw}=10^{-25}$, 100 bits for mantissa of real number and $\eps_m=1.57772\cdot 10^{-30}$. With such parameters of the method, the approximate values $x_1(T_c)$, $x_2(T_c)$ and $x_3(T_c)$ were also verified by the same numerical method, but in backward time. The values at the backward time in the end of the trajectory arc coincide with \reff{initvalcycle1} up to the 9th character inclusive after the point. The resulting values of $x_1(T_c)$, $x_2(T_c)$ and $x_3(T_c)$ coincide with \reff{initvalcycle1} up to the 8th character inclusive.

\begin{figure}[H]
 \includegraphics[width=\textwidth,height=10cm]{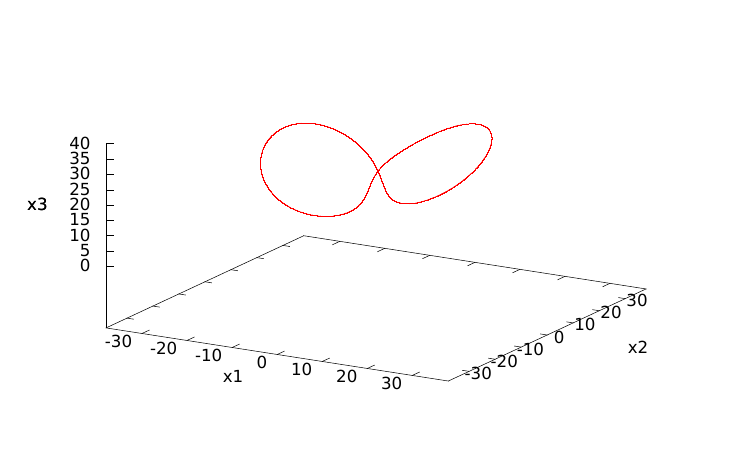}
 \caption{The cycle in the system \reff{lorenzsys} obtained by the FGBFI method.}\label{fig:1}
\end{figure}

The cycle corresponding to \reff{initvalcycle1} is shown in Figure \ref{fig:1}.

D. Viswanath \cite{Viswanath} found in the system \reff{lorenzsys} \textit{an unstable cycle} with a long period and with 99 decimal places use the Lindstedt--Poincar\'e technique:
\begingroup\makeatletter\def\f@size{8}\check@mathfonts
\equation
\begin{array}{c}
	x_1(0)\approx
	-13.568317317591138693791116532738086146665425413802770267307341928920639925115986035379124247913350182,\nl
	x_2(0)\approx
	-19.134575113926898610482106145675590555238063694018831440659257986585209042730623744601562225619287641,\nl
	x_3(0)= 27,\nl
	T_c\approx
	171.86372913973174676014481271369986804353836527546572814842984169003209300163561597123596376874805444.
\end{array}\label{initvalcycle2}
\endequation
\endgroup

The author of this article decided to check the found values with high accuracy in the forward time by the FGBFI method. For this, it was taken 390 bits for mantissa of real number. Then $\eps_m=7.93107\cdot 10^{-118}$. The value of accuracy for the power series is $\eps_{pw}=10^{-110}$. The maximum degree of the approximating polynomial at variable integration steps is 78. The computation time was approximately 6.2 min.

After the computational experiment, it was found that at the end of the period the values $x_1(T_c)$, $x_2(T_c)$ and $x_3(T_c)$ coincide up to 38 decimal places with \reff{initvalcycle2}. At the same time, the increase in accuracy (decreasing values $\eps_m$ and $\eps_{pw}$) does not affect the obtained results where we can conclude that some decimal places of the values \reff{initvalcycle2} are incorrect \mbox{in \cite{Viswanath}}, or need more correct decimal places.

Note that the classical numerical methods (such as the Euler method, Runge--Kutta methods, Adams methods, etc.) do not make it possible to carry out such a high-precision check of the values \reff{initvalcycle2} due to the instability of the found cycle by D. Viswanath.

We also note that the use of high-precision calculations is important in data encryption. For example, the results obtained in \cite{Flores} show that when using the high-precision arithmetic, the generated sequences provide good randomness and security with more iterations of the implemented discrete-time chaotic systems compared to the received sequences in the IEEE 754 format.

\section{The Systems of Explosive Type and Application of the FGBFI Method}
\label{sec:4}

\textit{The dynamic system of the explosive type} is a system for which the state $\|X(t)\|\rightarrow\infty$ is reached in a finite time. The examples of such systems are the system \reff{tan}, and the population explosion model
$$
 \dot{x}_1=kx_1^2,
$$
where the coefficient $k>0$.

In other words, in this section of the article we will consider the autonomous systems of differential equations whose solutions have the vertical asymptotes.
Such systems have been little-studied in the literature. We note the papers
\cite{Zhukovskii,Molokanova} where the scalar case is described: the asymptote existence theorem is proved and a modification of the Euler method is given which allows one to localize the asymptotes of solutions without \mbox{jumping them. }

For the multidimensional case, no studies have been found in the known literature. Most likely, this is due to the fact that for these systems, at $\|X(t)\|\rightarrow\infty$, the energy also tends to infinity (which cannot be in real systems according to the energy conservation law); however, the systems of explosive type can be models of real systems in some approximation.
Therefore, it is important for the researcher to have an idea at what time $t_{as}$ a sharp increase of phase coordinates will occur when it is not possible to find solutions of the system in quadratures.

Before proving the theorem on the existence of vertical asymptotes, we prove the following statement:

\begin{Lemma}
Let the number sequence $r^{(1)}$, $r^{(2)}$, ... converge to some number $r$, the sequence $w^{(1)}$, $w^{(2)}$, ... converge to number $w$, and there is the number $d>0$ that
\equation
 r^{(m)}-w^{(m)}\ge d\label{ineqseq1}
\endequation
for all $m\in\mathbb N$, then $r>w$.
\end{Lemma}

\begin{proof}
Let us represent the sequences as
$$
 \begin{array}{c}
 r^{(m)}=r+\zeta^{(m)},\nl
 w^{(m)}=w+\vartheta^{(m)},
 \end{array}
$$
where $\zeta^{(m)}$ and $\vartheta^{(m)}$ are the infinitesimal sequences.

From the inequality \reff{ineqseq1} it follows that
\equation
 r^{(m)}-w^{(m)}=(r-w)+\left(\zeta^{(m)}-\vartheta^{(m)}\right)\ge d>0.\label{ineqseq2}
\endequation

Note that the sequence
$$
 \zeta^{(m)}-\vartheta^{(m)}
$$
is also infinitesimal. Therefore, there exists a number $m=m^*$ such that for all $m>m^*$
$$
 \zeta^{(m)}-\vartheta^{(m)} \le \eps,
$$
moreover, we choose the small positive number $\eps$ such that
$$
 d-\eps>0,
$$
because the number $d$ is fixed. Then from \reff{ineqseq2}
$$
 (r-w)+\eps \ge d
$$
and
$$
 r-w\ge d-\eps>0,
$$
whence it follows that $r>w$.
\end{proof}

Note that the proof of this lemma can be carried over to the case when the sequences
$\left\{r^{(m)}(t)\right\}$ and $\left\{w^{(m)}(t)\right\}$ are the sequences of scalar functions that converge uniformly over some time interval.

Let us now prove a theorem that allows us to establish the existence of the vertical asymptotes in the system \reff{polsys} at $t>0$.

\begin{Theorem}
If some equation with the number $q$ of the system \reff{polsys} can be represented as ($k$ is some number of the phase coordinate $x_k$, $k\neq q$)
\equation
 \dot{x}_q=c_q+\langle \tilde{Q}_q (\tilde{X}-A_q),\tilde{X}-A_q \rangle+
 \upsilon x_k^2,\label{modifeq}
\endequation
where the symmetric matrix $\tilde{Q}_q$ is positive definite $(D_{\tilde{Q}}^+)$ or negative definite $(D_{\tilde{Q}}^-)$ and is obtained by deleting a row and a column with number $k$ from matrix $Q_q$,
$$
 \tilde{X}(t)=\left[x_1(t)\:\:\ldots\:\:x_{k-1}(t)\:\:x_{k+1}(t)\:\:\ldots\:\:x_{n}(t)\right]\tr,
 \:\:\tilde{X}\in\mathbb R^{n-1},
$$
$$
 A_q=\left[a_{q,1}\:\:\ldots\:\:a_{q,k-1}\:\:a_{q,k+1}\:\:\ldots\:\:a_{q,n}\right]\tr,
$$
$$
 \mbox{the numbers}\:\:\upsilon>0,\:\:c_q\ge0\:\:\mbox{for}\:\:D_Q^+
$$
or
$$
 \mbox{the numbers}\:\:\upsilon<0,\:\:c_q\le0\:\:\mbox{for}\:\:D_Q^-,
$$
the initial conditions $x_k(0)\neq0$, $x_q(0)>a_{q,q}$ for $D_Q^+$ or $x_q(0)<a_{q,q}$ for $D_Q^-$,
then the $q$-th component of the solution of the \mbox{system \reff{polsys}} has a vertical asymptote at $t>0$.
\end{Theorem}

\begin{proof}
For definiteness, we will consider the situation $D_{\tilde{Q}}^+$. A similar proof can be carried out for $D_{\tilde{Q}}^-$. Then the inequality (\cite{Demidovich}, pp. 34, 35)
$$
 \langle \tilde{Q}_q (\tilde{X}-A_q),\tilde{X}-A_q \rangle\ge\lambda_q\|\tilde{X}-A_q\|_2^2
$$
is valid where $\lambda_q>0$ is the smallest eigenvalue of the matrix $\tilde{Q}_q$. Since
$$
 \|\tilde{X}-A_q\|_2^2=\disp\sum_{\substack{p=1\\p\neq k}}^n(x_p-a_{q,p})^2\ge(x_q-a_{q,q})^2,
$$
then
\equation
 c_q+\langle \tilde{Q}_q (\tilde{X}-A_q),\tilde{X}-A_q \rangle\ge c_q+\lambda_q(x_q-a_{q,q})^2.
 \label{ineq2}
\endequation

Let us introduce the equation
\equation
 \dot{y}=c_q+\lambda_q(y-a_{q,q})^2\label{c1}
\endequation
with the initial condition $y(0)$ for which
\equation
 x_q(0)-a_{q,q}>y(0)-a_{q,q}>0.\label{c2}
\endequation

By the existence and uniqueness theorem, the solution of the Equations \reff{modifeq} and \reff{c1} exists and is unique on some time interval $[0,t^*]$.

Let us construct the integral equations equivalent to the Equations \reff{modifeq} and \reff{c1}:
\equation
 \begin{array}{c}
 \disp x_q(t)=x_q(0)+\int_0^t\left(c_q+
 \langle \tilde{Q}_q (\tilde{X}(\eta)-A_q),\tilde{X}(\eta)-A_q \rangle+
 \upsilon x_k^2(\eta)\right)d\eta,\nl
 \disp y(t)=y(0)+\int_0^t\left(c_q+\lambda_q(y(\eta)-a_{q,q})^2\right)d\eta.
 \end{array}\label{int_2}
\endequation

Since the function $x_k(t)$ is continuous on the interval $[0,t^*]$, $x_k(0)\neq0$ and $\upsilon>0$, then
$$
 R(t)=\disp\upsilon\int_0^t x_k^2(\eta)d\eta>0
$$
for $t\in(0,t^*]$. Let $R^*=R(t^*)$ is a maximum value of the function $R(t)$ on the segment $[0,t^*]$.

Let us write the function $x_q(t)$ as
$$
 \disp x_q(t)=R(t)+x_q(0)+\int_0^t\left(c_q+
 \langle \tilde{Q}_q (\tilde{X}(\eta)-A_q),\tilde{X}(\eta)-A_q \rangle\right)d\eta.
$$

Let us now consider the schemes of successive approximations obtained from the Formula \reff{int_2}:
$$
 \begin{array}{c}
 \disp x_q^{(m+1)}(t)=R(t)+x_q^{(m)}(0)+\int_0^t\left(c_q+
 \langle \tilde{Q}_q (\tilde{X}^{(m)}(\eta)-A_q),\tilde{X}^{(m)}(\eta)-A_q \rangle
 \right)d\eta,\nl
 \disp y^{(m+1)}(t)=y^{(m)}(0)+\int_0^t\left(c_q+\lambda_q(y^{(m)}(\eta)-a_{q,q})^2\right)d\eta,
 \end{array}
$$
where the vector function $\tilde{X}^{(m)}(\eta)$ is obtained from $\tilde{X}(\eta)$ by replacing the function $x_q(\eta)$ by $x_q^{(m)}(\eta)$. Let
$$
 \begin{array}{c}
 x_q^{(0)}(t)\equiv x_q(0),\nl
 y^{(0)}(t)\equiv y(0).
 \end{array}
$$

Since the existence and uniqueness theorem is proved by the method of successive approximations, then for the constructed approximations
$$
 \begin{array}{c}
 \disp\lim_{m\rightarrow\infty}x_q^{(m)}(t)=x_q(t),\nl
 \disp\lim_{m\rightarrow\infty}y^{(m)}(t)=y(t)
 \end{array}
$$
uniformly on the time interval $[0,t^*]$.

By virtue of the inequality \reff{ineq2}, at the first iteration we have
$$
 \disp x_q^{(1)}(t)\ge R(t)+x_q(0)+\int_0^t\left(c_q+\lambda_q(x_q(0)-a_{q,q})^2\right)d\eta.
$$

From the positiveness of the function $R(t)$, the inequalities \reff{c2} and
$$
 (x_q(0)-a_{q,q})^2>(y(0)-a_{q,q})^2,
$$
we have
\equation
 x_q^{(1)}(t)>y(0)+\int_0^t\left(c_q+\lambda_q(y(0)-a_{q,q})^2\right)d\eta\equiv y^{(1)}(t),\label{ineqx}
\endequation
i.e., $x_q^{(1)}(t)>y^{(1)}(t)$ for $t\in[0,t^*]$.

Also from the the inequalities \reff{c2} and \reff{ineqx}, we have
$$
 x_q^{(1)}(t)-a_{q,q}>y^{(1)}(t)-a_{q,q}>0.
$$

Let
$$
 x_q^{(m)}(t)-a_{q,q}>y^{(m)}(t)-a_{q,q}>0
$$
for $t\in[0,t^*]$.
Hence we get
$$
 x_q^{(m)}(t)>y^{(m)}(t).
$$
In a similar way, one can show that
$$
 x_q^{(m+1)}(t)-a_{q,q}>y^{(m+1)}(t)-a_{q,q}>0
$$
and
$$
 x_q^{(m+1)}(t)>y^{(m+1)}(t).
$$
At that
$$
 \max_{t\in[0,t^*]}\left(x_q^{(m+1)}(t)-y^{(m+1)}(t)\right)\ge R^*>0
$$
for all $m\in\mathbb N$.

Then, by virtue of the proved lemma
\equation
 x_q(t)>y(t)\label{ineq3}
\endequation
at $t\in[0,t^*]$.

Note that the solution to the Cauchy problem \reff{c1}--\reff{c2}, which is easy to find in quadratures, has a vertical asymptote. Then it follows from the inequality \reff{ineq3} that the function $x_q(t)$ also has a vertical asymptote.
\end{proof}

Let us consider an example of a system that satisfies the conditions of this theorem.
In the article \cite{Gotthans}, the following system was introduced:
\equation
 \left\{\begin{array}{l}
 \dot{x}_1=ax_3,\nl
 \dot{x}_2=x_3f_1(x_1,x_2,x_3),\nl
 \dot{x}_3=x_1^2+x_2^2-r^2+x_3f_2(x_1,x_2,x_3),\label{newsysgen}
 \end{array}\right.
\endequation
in which the equilibrium positions lie on the circle $x_1^2+x_2^2=r^2$ in the plane $x_3=0$, $a$ is the parameter of the dynamical system. Usually the functions $f_1$ and $f_2$ are chosen as
$$
 \begin{array}{c}
 f_1(x_1,x_2,x_3)=bx_1+\beta x_3^2,\nl
 f_2(x_1,x_2,x_3)=\gamma x_1,
 \end{array}
$$
where $b$, $\beta$ and $\gamma$ are also system parameters.

Let $r=0$, $\beta=0$ and
$$
 f_2(x_1,x_2,x_3)=\alpha x_3,
$$
where $\alpha>0$ is a parameter. We get
\equation
 \left\{\begin{array}{l}
 \dot{x}_1=ax_3,\nl
 \dot{x}_2=bx_1x_3,\nl
 \dot{x}_3=x_1^2+x_2^2+\alpha x_3^2.
 \end{array}\right.\label{newsys}
\endequation

The system \reff{newsys} belongs to the class of systems \reff{polsys} and has the unique equilibrium $(0,0,0)$. For the last equation of the system
$$
 c_3=0,\:\:\tilde{Q}_3=\left[
 \begin{array}{rrr}
 1 & 0\nl
 0 & \alpha
 \end{array}
 \right],\:\:A_3=\nulmt,
$$
the matrix $\tilde{Q}_3$ is symmetric and positive--definite. Then the component $x_3(t)$ of the solution of the system \reff{newsys} has a vertical asymptote for $x_3(0)>0$ and $x_1(0)\neq0$.

Note that the authors of the article \cite{Gotthans} do not provide a qualitative analysis of the system \reff{newsysgen}. As shown above, it may happen that it has unbounded solutions. Therefore, before looking numerically for attractors in dynamical systems, the researcher needs to show the existence of bounded solutions, for example, using the Lyapunov functions.

Note that by the proof of the above theorem, it is not necessary that all equations of the system \reff{polsys} have a quadratic right-hand side. It suffices to have at least one equation that satisfies the conditions of this theorem.

It follows from the Formula \reff{tauc} that the application of the FGBFI method allows one to approach the vertical asymptote arbitrarily close without jumping over it, since
$$
 \tau(\lam_0)=O\left(\dfrac{1}{\|\lam_0\|^2}\right)
$$
at $\|\lam_0\|\rightarrow\infty$,
i.e., the Formula \reff{tauc} give a guaranteed region of convergence of the power series, and for larger $\|\lam_0\|$ the integration step is smaller.
Thus, from step to step applying Algorithm \ref{alg:FGBFI}, we will come as arbitrarily close to time $t_{as}$ localizing the moment of explosion.

\textit{Note also that the application of the FGBFI method for the numerical integration of the equation \reff{tan} gives an approximate value of the number $\pi$ localizing the asymptote \reff{astan}.}

\section{The Piecewise Smooth Systems with Quadratic-Type Nonlinearities. The Systems with Hysteresis}
\label{sec:5}

In practice, the problem often arises of studying piecewise smooth systems, for example, the Chua system \cite{Kuznetsov}:
\equation
 \left\{\begin{array}{l}
 \dot{x}_1=\rho(x_2-(m_1+1)x_1)-\rho\phi(x_1),\nl
 \dot{x}_2=x_1-x_2+x_3,\nl
 \dot{x}_3=-(\theta x_2+\chi x_3),
 \end{array}\right.\label{chua}
\endequation
where $\rho$, $\theta$, $\chi$, $m_0$ and $m_1$ are system parameters, the function
\equation
 \phi(x_1)=\dfrac{1}{2}(m_0-m_1)(|x_1+1|-|x_1-1|).\label{phix}
\endequation

The authors of the article \cite{Kuznetsov} use the method of a describing function to localize hidden attractors in the system \reff{chua}, and the values of the initial conditions are also given. Therefore, the task of developing the high-precision numerical methods for constructing approximate solutions to systems of the form \reff{chua} is important. Note that the described idea can also be transferred to models with hysteresis which were considered in \cite{MedvedskiiMeleshenko,MeleshenkoSemenov,SemenovReshetova}.

We can extend the FGBFI method to piecewise smooth systems with quadratic nonlinearities. Let us describe the idea of the method using the system \reff{chua} as an example. 

From the form of the function $\phi(x_1)$, the phase space is divided into three parts, where the right side of the \reff{chua} system is smooth. Similarly, the phase space can be divided into regions of smoothness for systems with hysteresis. For example, for systems with a non-smooth potential \cite{MeleshenkoSemenov}. The joint point coordinates are
\equation
 x_1=\pm1.\label{pl}
\endequation

By the initial condition for the coordinate $x_1$, we determine in which part of the phase space the initial point is located.
The modules in the Formula \reff{phix} expand accordingly. Next, we take a forward step in time $\Delta t$ according by Algorithm \ref{alg:FGBFI}.
In this case, it is necessary to remember the obtained polynomials $\tilde{x}_1(t)$, $\tilde{x}_2(t)$ and $\tilde{x}_3(t)$ approximating the corresponding phase coordinates $x_1(t)$, $x_2(t)$ and $x_3(t)$ on the time interval $[0,\Delta t]$. If $\tilde{x}_1(\Delta t)$ does not belong to the current part of the phase space, then it is necessary to find the time $t_{pl}$ with a high accuracy when the trajectory intersects one of the planes \reff{pl}.
For this, for example, one of the equations
$$
 \tilde{x}_1(t_{pl})=\pm1
$$
is numerically solved by the secant method.

Further, the vector of initial conditions is taken equal to
$$
 \lam_0=\left[\tilde{x}_1(t_{pl})\:\:\tilde{x}_2(t_{pl})\:\:\tilde{x}_3(t_{pl})\right]\tr,
$$
and for the new part of the phase space, the modules are again expanded in the \mbox{Formula \reff{phix}}.

The advantage of the described algorithm is that we find the moments of intersection of the trajectory of the system \reff{chua} with the planes \reff{pl} with a high accuracy, slightly accumulating calculation errors.

\section{Conclusions}

In this article, the high-precision FGBFI method for constructing approximations to unstable solutions of dynamic systems of the form \reff{polsys} was described, and a link to the developed program software was also provided. The arbitrary-precision numbers were represented by \textit{mpreal} data type of the MPFR C++ library with overloaded arithmetic operations, as well as friendly mathematical functions.

For systems of explosive type, a theorem is proved that allows one to establish the existence of asymptotes in the system \reff{polsys}, and the advantage of the numerical FGBFI method in this case is also shown.

The FGBFI method has also been extended to piecewise smooth systems with quadratic-type nonlinearities using the Chua system as an example, as well as systems with hysteresis.

\reftitle{References}

\end{document}